\documentstyle[12pt,epsf]{article} 

\newcommand{\bi}{\begin{itemize}}
\newcommand{\ei}{\end{itemize}}
\newcommand{\bc}{\begin{center}}
\newcommand{\ec}{\end{center}}

\newcommand{\be}{\begin{equation}}
\newcommand{\ee}{\end{equation}}

\newcommand{\bqn}{\begin{eqnarray}}
\newcommand{\eqn}{\end{eqnarray}}

\begin{document}

\title{Quantum Properties of Solitons in $U(\phi)=\phi^2\ln^2(\phi^2)$
and $U(\phi)=\phi^2\cos^2(\phi^2)$ Models} 

\author{Gabriel H. Flores$^{1}\;$\thanks{E-mail:
gflores@lafex.cbpf.br}$\;$,and 
N.F. Svaiter.$^{2}\;$\thanks{Permanent address:
Centro Brasileiro de Pesquisas Fisicas, Rio De Janeiro, RJ 22290-180,
Brazil. E-mail: svaiter@lns.mit.edu, nfuxsvai@cbpf.br}\\ \\
{\it $^1\;$ Centro Brasileiro de Pesquisas Fisicas-CBPF,}\\
{\it Rua Dr.Xavier Sigaud 150, 22290-180 Rio de Janeiro, RJ, Brazil} 
\\
{\it $^2 \;$ Center of Theoretical Physics,}\\
{\it Massachusetts Institute of Technology,}\\
{\it Cambridge, MA 02139, USA.}}

\maketitle

\begin{abstract} 
 Recently we constructed two new $(1+1)$-dimensional scalar
field theory models that posses solitonic solutions. They are
the $U(\phi)=\phi^2\ln^2(\phi^2)$ and the 
$U(\phi)=\phi^2\cos^2(\phi^2)$ models . The first quantum 
corrections for these models are given by exactly solvable Schrodinger equations. In this paper we first  examine the quantum meaning of the solitonic solutions and study the scattering of the mesons by the 
quantum soliton at order $\hbar$. Finally we give a finite expression for the  
soliton masses of both models and evaluate such expression approximately in the case of the second model.
\vspace{0.34cm} 
\noindent \\
PACS number(s):11.30.Pb, 03.65.Fd, 11.10.Ef.    
\end{abstract} 

\newpage

\section{Introduction}
\begin{figure}
\centerline {\epsfxsize=3in\epsfysize=1.5in\epsffile{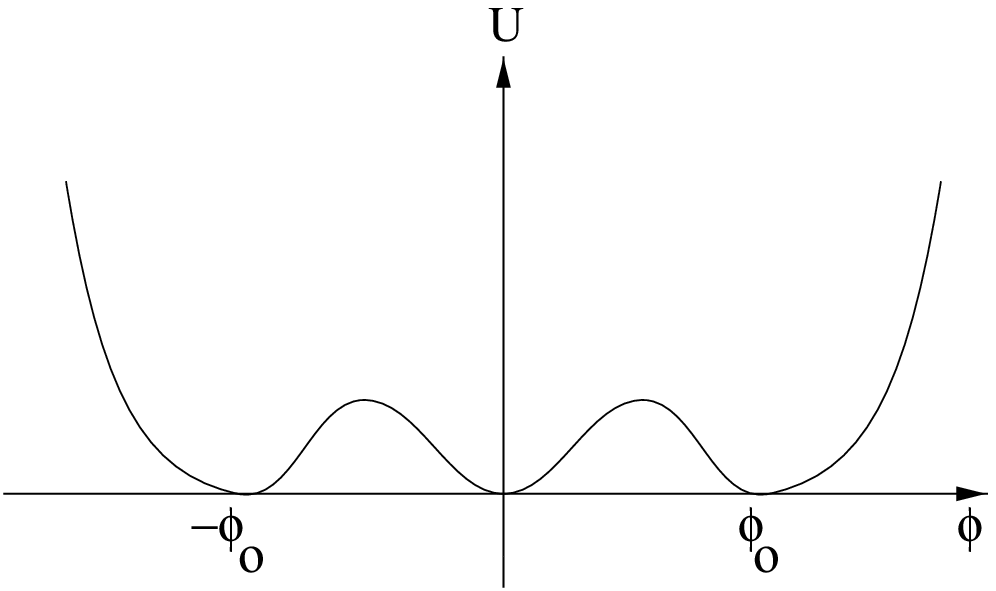} }
\caption{The density potential $U(\phi)$ given by eq.(\ref{new1}).}
\begin{picture}(10,10)
\end{picture}
\label{morsef}
\end{figure}

\begin{figure}
\centerline {\epsfxsize=3.2in\epsfysize=1.8in\epsffile{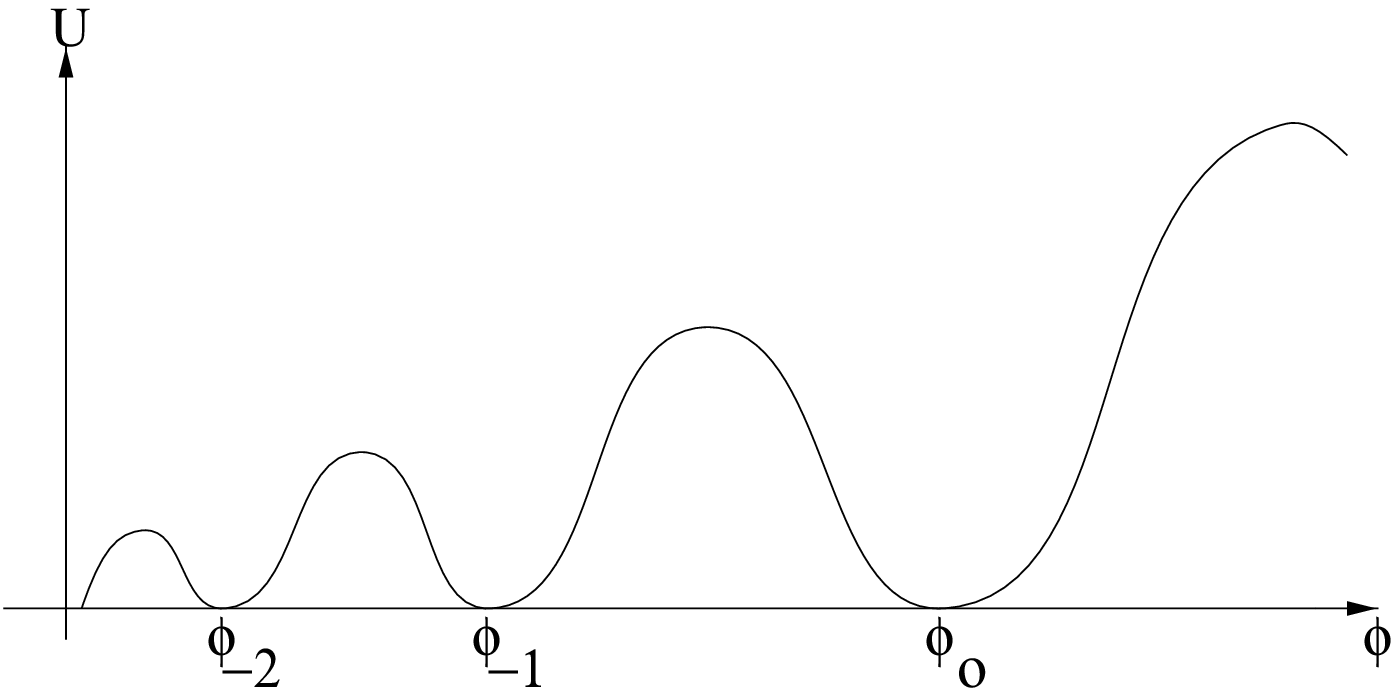} }
\caption{The density potential $U(\phi)$ given by eq.(\ref{new2}).}
\begin{picture}(10,10)
\end{picture}
\label{scarff}
\end{figure}

In a recent work \cite{flores1} the authors have constructed two new
$(1+1)$-dimensional scalar field theory models starting from analyticaly solvable
Schrodinger equations. Starting from the Morse potential we have
obtained the model with density potential given by
\be
U(\phi)=\frac{m^2}{8}\phi^2\ln^2\left(\frac{\alpha^2\phi^2}
{9m^4}\right)\;,
\label{new1}
\ee
and starting from the Scarf II hyperbolic potential we obtained the model
with density potential given by
\be
U(\phi)=\frac{1}{2}m^2B^2\phi^2\cos^2 \left[ \frac{1}{2B}\ln\left(\frac{\alpha^2\phi^2}{9m^4}\right)\right]\;.
\label{new2}
\ee
(In ref.\cite{flores1} we fixed B=1/2) The profiles of these density potentials given by eqs. (\ref{new1}) and
(\ref{new2}) are showed in fig.(\ref{morsef}) and fig.(\ref{scarff})
respectively. In that follows we refer to the models with density 
potentials given by equations (\ref{new1}) and (\ref{new2}) as models I and 
II respectively. 

For the model I it is easy to obtain the kink and anti-kink like solutions. 
They are given by 
\be
\phi_c(x)=\pm\phi_0\exp(-e^{\pm m(x-x_0)})\;,
\label{kink1}
\ee
where $\phi_0=3m^2/\alpha$ \cite{flores1}. In this case we have two kinks and two anti-kinks respectively that
lie between two of the three degenerate trivial vacua that are showed
in fig.(\ref{morsef}). The energy of these classical configurations are
all the same and are given by
\bqn
H[\phi_c]&=&\int_{-\infty}^{\infty}dx\left (\frac{1}{2}\left(\frac{d\phi_c}{dx}\right)^2+
\frac{m^2}{8}\phi_c^2\left[\ln\left(\frac{\alpha^2\phi_c^2}
{9m^4}\right)\right]^2\right)\nonumber\\
&=&\frac{9m^6}{2\alpha^2}\int_{-\infty}^{\infty}dxe^{\pm mx}\exp(-2e^{\pm mx})
\nonumber\\
&=&\frac{9m^5}{4\alpha^2}\;.
\label{enrc1}
\eqn
We can make a Taylor expansion around one of the trivial vacua and
we obtain 
\be
U(\varphi)=\frac{1}{2}m^2\varphi^2\pm\frac{\alpha}{6}\varphi^3-
\frac{1}{216}\varphi^4+\frac{1}{14580}\frac{\alpha^4}{m^6}\varphi^6+...
\label{taylor1}
\ee
where we have shifted the field as $\phi=(\varphi\pm\phi_0)$. 
We remark that
an expansion around the point $\phi=0$ is not possible because with the
exception of the first derivative all
the derivatives of $U(\phi)$  are infinite at this point. This means that
it is not possible to have perturbative dynamics around the trivial vacuum $\phi=0$. 
Note that this model possesses a discrete symmetry $U(\phi)=U(-\phi)$ that
is broken by the perturbative vacua $\pm \phi_0$. 

Model II has infinitely degenerate trivial vacua at the points
$\phi=\pm\phi_n$ with $\phi_n$ given by
\be
\phi_n=\frac{3m^2}{\alpha}\exp\left(\frac{2n+1}{2}\pi B\right)\;, ~~n=0,
\pm 1,\pm 2,..
\label{zeros2}
\ee
The kinks and anti-kinks that lie between two of these trivial vacua are given by
\be
\phi_c(x)=\pm\frac{3m^2}{\alpha}\exp\left(n\pi B
 \pm B\tan^{-1}(\sinh(mx))\right)\;,~~n=0,
\pm 1,\pm 2,..
\label{kink2}
\ee
where the solutions with $(\pm)$ signs in the exponents correspond to the 
kinks and anti-kinks solutions respectively for each value of $n$ and for each 
sign that appears in front. Also this model possesses the discrete symetry $U(-\phi)=U(\phi)$ that is broken by the trivial vacua.    
Notice that $\lim_{n\rightarrow\infty}\phi_n=0$, {\it i.e}, the density potential oscillates infinite times between some finite value of $\phi$ and zero.  Also observe that the distance between the trivial vacua grows out  exponentially. We can also obtain the energy for these configurations,
they are given by
\bqn
H[\phi_c]&=&\int_{-\infty}^{\infty}\left(
\frac{1}{2}\left(\frac{d\phi_c}{dx}\right)^2+
\frac{m^2}{8}\cos^2\left[\ln\left(\frac{\alpha^2\phi^2}{9m^4}\right)\right]
\right)
\nonumber\\
&=&\frac{9m^6B^2e^{2Bn\pi}}{\alpha^2}\int_{-\infty}^{\infty}dx
\frac{\exp(\pm 2B\arctan(\sinh(mx)))}{\cosh^2(mx)}
\nonumber\\
&=&\frac{9m^5B^2e^{2Bn\pi}}{\alpha^2}I(B)
\label{enrc2}
\eqn
where $I(B)$ is given by
\be
I(B)=\int_{-1}^{1}ds e^{2B\arcsin(s)}
\label{numer}
\ee
and can be evaluated numericaly for each value of $B$. In eq. (\ref{enrc2}) we can
see that the energy associated with these static configurations
are all different. As in the case of the model I, we can make a Taylor expansion
around one of the trivial vacua and we obtain
\bqn
U(\varphi)&=&\frac{m^2}{2} \varphi^2 
\pm\frac{\alpha}{6}e^{-\frac{1}{2}(2n+1)\pi B} \varphi^3
-\frac{(1+4/B^2)}{9(4!)}\frac{\alpha^2}{m^2}e^{-(2n+1)\pi B}\varphi^4
\nonumber\\
& &~~~~~~~~~~~~~~~~~~~~~~
+\frac{(4+20/B^2+16/B^4)}{(81)6!}\frac{\alpha^4}{m^6}e^{-2(2n+1)\pi B}\varphi^6+...
\label{taylor2}
\eqn
Here we have redefined $\phi=(\varphi\pm\phi_n)$. From the Taylor expansion
we can see a peculiarity of this model. When $n\rightarrow\infty$ all
the terms vanish except the quadratic term, {\it i.e}, the perturbative sector around
the trivial vacuum for $n\rightarrow\infty$ becomes a free theory. Also as in the model
I we have to stress
that when $n\rightarrow-\infty$ the Taylor expansion given by 
eq.(\ref{taylor2}) is meaningless, {\it i.e}, 
there is no perturbative dynamics around the trivial vacuum $\phi=0$. 

In the next section following Jackiw\cite{jackiw} we will use the
Kerman-Klein  method to analyse the quantum properties of the solitonic
solutions. 

\section{Quantum meaning of the solitonic solutions: the Kerman-Klein method} 
To analyse the quantum
properties of the solitonic solutions, we use the Kerman-Klein 
method \cite{jackiw1}, \cite{jackiw2}. For such purpose we postulate
that in addition to the perturbative states (that we call meson states), there
are other particle states, the quantum soliton states
(in the literature they are called the baryons states). The one-soliton state
$|E>$ are energy and momentum eigenstates
\bqn
\hat{H}|E>=E|E>\;,\nonumber\\
\hat{P}|E>=P|E>\;,
\eqn
where $E=\sqrt{P^2+M^2}$ and $M$ is the mass of the quantum soliton state.
Further it is postulated that for small coupling constant ($\alpha$ small)
the quantum soliton is very heavy, {\it i.e}, $M$ is taken to be
\be
M\approx{\mathcal O}(\alpha^{-2})\;.
\label{post1}
\ee
In addition we postulate that there are one-soliton multi-meson states,
$|P,K_1,K_2...>$ where $P$ is the total momentum and $k_i$ is the asymptotic
meson momentum. Also we postulate the stability of the quantum solitonic 
state, {\it i.e}, that matrix elements as
\be
<{\mathrm meson~state,~soliton~state}|\hat{\phi}\hat{\phi}...|
{\mathrm meson ~state, ~no ~soliton~ state}>=0
\ee
vanish. In model I we can see that there are two types of soliton
states, {\it i.e}, the corresponding to the ($\pm$) signs that appear in
the exponential of eq.(\ref{kink1}). These states are called soliton
and anti-solitons states respectively. For the model II we have an infinite
number of soliton and anti-soliton states, labeled by the index $n$
of eq. (\ref{kink2}). In both cases we are considering that the soliton
and anti-soliton states corresponding to the ($\pm$) signs that appear in front
to the eqs. (\ref{kink1}) and (\ref{kink2}) are the same. This is justified
by the discrete symetry $U(\phi)=U(-\phi)$ in both cases.
The sectors of the Hilbert state constructed taking into acount the
soliton and anti-soliton states are called the soliton and anti-soliton 
sectors respectively. Finally we postulate that the soliton sectors do not
comunicate with the anti-solitonic sectors. Now we can verify using the
above postulates and self-consistence that
\be
<P';K'_1,...,K'_n|\hat{\phi}|P;K_1,...,K'_m>_{c}~\!\approx
{\mathcal O}(\alpha^{n+m-1})\;,
\label{self}
\ee
where the subscript $c$ denotes connected part and $\hat\phi$ is a
Heisenberg field operator. For this purpose we
use the equation of motion for the field operator $\phi$,
\be
\frac{\partial^2\hat{\phi}} {\partial t^2}-
\frac{\partial^2\hat{\phi}}{\partial x^2}=-U'(\hat{\phi})\;.
\label{fieldeq}
\ee
In first place we will compute $f(P',P)=<P'|\hat{\phi}|P>$ to the
order ${\mathcal O}(\alpha^{-1})$. In both models we can make a
taylor expansion for $U(\hat{\phi})$ around one of the trivial vacua. It is easy
to show that such expression in both cases have the following form
\be
U(\hat{\phi})=\sum_{l=2}^{\infty}a_l\alpha^{l-2}(\hat{\phi}-\phi_0)^{l}\;.
\label{taylorp}
\ee
In eq.(\ref{taylorp}) $\phi_0$ denotes one of the trivial vacua and 
the coefficient $a_l$ does not depend on $\alpha$. Using eq.(\ref{taylorp}) in
eq.(\ref{fieldeq}) we have
\be
\frac{\partial^2\hat{\phi}}{\partial t^2}-
\frac{\partial^2\hat{\phi}}{\partial x^2}=-\sum_{l=2}^{\infty}la_l\alpha^{l-2}
(\hat{\phi}-\phi_0)^{l-1} \;.
\label{taylor3}
\ee
Using the Heisenberg equations of motion
\bqn
\frac{\partial \hat{\phi}}{\partial t}=i[\hat{H},\hat{\phi}]\nonumber\\
\frac{\partial \hat{\phi}} {\partial x}=-i[\hat{P},\hat{\phi}]
\label{heisem}
\eqn
twice in eq. (\ref{taylor3}) we obtain
\be
\left((P-P')^2-(E-E')^2\right)f(P',P)=-\sum_{l=2}^{\infty}la_l\alpha^{l-2}
<P'|(\hat{\phi}-\phi_0)^{l-1}|P>\;.
\label{form}
\ee
Each element $f^{l-1}(P',P)=<P'|(\hat{\phi}-\phi_0)^{l-1}|P>$ can be written as
\bqn
f^{l-1}(P',P)&=&
\sum_{m_i}
<P'|(\hat{\phi}-\phi_0)|m_1>
<m_1|(\hat{\phi}-\phi_0)|m_2>
<m_2|...\nonumber\\
& &~~~~~~~~~~~~~...|m_{l-2}><m_{l-2}|(\hat{\phi}-\phi_0)|P>
\;,
\label{comple1}
\eqn
where the sum is over the complete basis $|m_i>$, {\it i.e}, over the
soliton states, the soliton one meson states, the soliton two meson states
, etc. In eq.(\ref{form}) the 
left hand side is of the  order ${\mathcal O}(\alpha^{-1})$. For 
consistency we have to take the right hand side also
of this order. In order to make this possible we have 
to take each $f^{l-1}(P',P)$
term of the order ${\mathcal O}(\alpha^{-(l-1)})$. 
Then in eq.(\ref{comple1}) we have to retain only the sum over the 
one solitonic (or one anti-solitonic, if it were the case) states.
Then we have for $f^{l-1}(P',P)$:
\bqn
& &\int \frac{dP_1}{2\pi}\frac{dP_2}{2\pi}...
\frac{dP_{l-2}}{2\pi}[f(P',P_1)-2\pi\phi_0\delta(P'-P_1)]
[f(P_1,P_2)-2\pi\phi_0\delta(P_1-P_2)]...
\nonumber\\
& &~~~~~~~~~~~~~~~~~~~~~~~~~~...[f(P_{l-2},P)-2\pi\phi_0\delta(P_{l-2}-P)]\;.
\label{comple2}
\eqn
Replacing eq.(\ref{comple2}) in eq.(\ref{form}) we obtain 
an integral equation for $f(P',P)$,
\bqn
\!\!\!\!\!\!\!\!\!\!\!\!\!\!\left( (P-P')^2-(E-E')^2 \right) f(P',P)=
-\sum_{l=2}^{\infty}la_l \alpha^{l-2}
\int \frac{dP_1}{2\pi} ...\frac{dP_{l-2}}{2\pi} \nonumber\\
~~~~~~~~~[f(P',P_1)-2\pi\phi_0\delta(P'-P_1)]
...[f(P_{l-2},P)-2\pi\phi_0\delta(P_{l-2}-P)]\;.
\label{integral}
\eqn
To solve eq.(\ref{integral}) we note that at this order 
$(E-E')\approx 0$. This can be seen
using $E=\sqrt{P^2+M^2}\approx M+\frac{P^2}{2M}$ and noting 
that the second term is of the order ${\mathcal O}(\alpha^2)$, 
{\it i.e}, at this order the energy is independent of
the momentum. This is compatible with the assumption that
the quantum soliton is very heavy. Also
using the Lorentz invariance of $f(P',P)$ we can set
at this order  $f(P',P)=f(P-P)$. Using 
these facts eq. (\ref{integral}) can be reduced to
\bqn
(P-P')^2f(P',P)&=&-\sum_{l=2}^{\infty}la_l\alpha^{l-2}
\int \frac{dP_1}{2\pi}...
\frac{dP_{l-2}}{2\pi}[f(P'-P_1)-2\pi\phi_0\delta(P'-P_1)]...\nonumber\\
& &~~~~~~~~~~~~~~~~~~~...
[f(P_{l-2}-P)-2\pi\phi_0\delta(P_{l-2}-P)]\;.
\label{four}
\eqn
We can solve eq.(\ref{four}) in terms of a Fourier transformed function $\phi(x)$
\be
f(P'-P)=\int dx \exp[i(P'-P)x]\phi(x)\;.
\label{fourier}
\ee
In term of $\phi(x)$ eq. (\ref{four}) is
\bqn
\phi''(x)&=&\sum_{l=2}^{\infty}la_l\alpha^{l-2}\left(\phi(x)-\phi_0\right)^{l-1}
\nonumber\\
&=&U'(\phi).
\label{classic}
\eqn
We recognize eq.$(\ref{classic})$ as the classical equation  of motion for 
$\phi_c$. Therefore, at the order ${\mathcal O}(\alpha^{-1})$, we obtain for the matrix 
elements for the quantum field operator between one soliton (or one anti-soliton) states:
\be
<P'|\hat{\phi}|P>=\int dx \exp[i(P'-P)x]\phi_c(x)\;,
\label{factor}
\ee
where $\phi_c(x)$ is given by equation (\ref{kink1}) or (\ref{kink2}). Since in both models
I and II $\phi_c(x)$ is proportional to $1/\alpha$ one verifies in eq.(\ref{factor})
that  $<P'|\hat{\phi}|P>$ is ${\mathcal O}(\alpha^{-1})$.
We can compute at first leading order the mass of the quantum solitonic 
state using that
at leading order $\hat{H}|P>=\sqrt{P^2+M^2}|P>\approx M|P>$. The energy
operator is
\be
\hat{H}=\int dx \left( \frac{1}{2} \partial_\mu \hat{\phi}\partial_\mu \hat{\phi}
+U(\hat{\phi}) \right)\;.
\label{energy}
\ee
Replacing eq.(\ref{taylorp}) in eq. (\ref{energy}) we obtain
\bqn
<P|\hat{H}|P>&=&\int dx \left( \frac{1}{2} 
<P|\partial_\mu \hat{\phi}\partial_\mu \hat{\phi}
|P>+\sum_{l=2}^{\infty}
a_l\alpha^{l-2}<P|(\hat{\phi}-\phi_0)^l|P>\right)\nonumber\\
&=&\int dx \left(\frac{1}{2}\sum_{m_1}<P|\partial_\mu \hat{\phi}
|m_1><m_1|\partial_\mu \hat{\phi}|P>
+\sum_{l=2}^{\infty}\sum_{m_i}a_l\alpha^{l-2}\right.\nonumber\\
& &\left.<P|(\hat{\phi}-\phi_0)|m_1><m_1|...|m_{l-1}><m_{l-1}|(\hat{\phi}-\phi_0)|P>
\right)
\label{sumad}
\eqn
 and since the left hand side of eq.(\ref{sumad}) is of ${\mathcal O}(\alpha^{-2})$
order we can see that in the sum over $m_i$ we have only to retain
the sum over the one soliton states and using eq.(\ref{heisem}) in the first 
term of the right hand side of eq.(\ref{sumad}) we obtain
\bqn
<P|\hat{H}|P>&\approx&2\pi\delta(P-P)M\nonumber\\
&=&2\pi\delta(0) \int dx\left(\frac{1}{2}\left(\frac{d\phi_c}{dx}\right)^2
+U(\phi_c)\right)\;.
\label{mass}
\eqn
Then we have for the soliton mass at leading order
\be
M= \int dx\left(\frac{1}{2}\left(\frac{d\phi_c}{dx}\right)^2
+U(\phi_c)\right)\;,
\label{masscl}
\ee
that is, it is equal to the energy of the classical static configuration.
For model I it is given by the eq.(\ref{enrc1}) and for model II it is
given by eq.(\ref{enrc2}). In both cases $M$ is ${\mathcal O}(\alpha^{-2})$
confirming eq.(\ref{post1}). Note that in model I the two different soliton
states have the same mass at this order. On the other hand the mass
of the soliton states of the model II are different. Note also that
the soliton mass becomes infinite when $n\rightarrow\infty$.
In order to compute $f(P',P)$ to the next order we have to retain
in eq.(\ref{comple1}) the sum over soliton one meson states $|P,k>$,
then we will obtain the next term for $f^{l-1}(P',P)$:
\bqn
\int\frac{dP_1}{2\pi}\frac{dk_1}{2\pi}\frac{dP_2}{2\pi}...
\frac{dP_i}{2\pi}...\frac{dP_{l-2}}{2\pi}
<P'|(\hat{\phi}-\phi_0)|P_1;k_1><P_1;k_1|(\hat{\phi}-\phi_0)|P_2><P_2|...
\nonumber\\
|P_i><P_i|(\hat{\phi}-\phi_0)|P_{i+1}><P_{i+1}|...
|P_{l-2}><P_{l-2}|(\hat{\phi}-\phi_0)|P>+...\nonumber\\
...+\int\frac{dP_1}{2\pi}\frac{dP_2}{2\pi}...
\frac{dP_i}{2\pi}\frac{dk_i}{2\pi}...\frac{dP_{l-2}}{2\pi}
<P'|(\hat{\phi}-\phi_0)|P_1><P_1|(\hat{\phi}-\phi_0)|P_2><P_2|...
\nonumber\\
|P_i;k_i><P_i;k_i|(\hat{\phi}-\phi_0)|P_{i+1}><P_{i+1}|...
|P_{l-2}><P_{l-2}|(\hat{\phi}-\phi_0)|P>+...\nonumber\\
...+\int\frac{dP_1}{2\pi}\frac{dP_2}{2\pi}\frac{dP_i}{2\pi}...
\frac{dP_{l-2}}{2\pi}\frac{dk_{l-2}}{2\pi}
<P'|(\hat{\phi}-\phi_0)|P_1><P_1|(\hat{\phi}-\phi_0)|P_2><P_2|...
\nonumber\\
|P_i><P_i|(\hat{\phi}-\phi_0)|P_{i+1}><P_{i+1}|...
|P_{l-2};k_{l-2}><P_{l-2};k_{l-2}|(\hat{\phi}-\phi_0)|P>.
\label{next}
\eqn
Since $f_k(P_1,P_2)=<P_1|\hat{\phi}|P_2,k>$ is of the order $O(\alpha^0)$
then the contribution for $f^{l-1}(P',P)$ given by eq.(\ref{next}) is of
the order $O(\alpha^{3-l})$ and from eq.(\ref{form}) we can see that
the expression (\ref{next}) is the $O(\alpha^1)$ contribution for $f(P',P)$.
We can see that to compute $f(P',P)$ to the next order we need to know $f_k(P_1,P_2)$ at lower order. Also if one computes the energy, to the next
order we will need $f_k(P_1,P_2)$. Then our next task is to evaluate
$f_k(P_1,P_2)$. Using eq.(\ref{taylor3}) and eq.(\ref{heisem}) we obtain
\be
\left((P_2-P_1)^2-(E_2+\omega(k)-E_1)^2\right)f_k(P_1,P_2)=
-\sum_{l=2}^{\infty}la_l\alpha^{l-2}
<P_1|(\hat{\phi}-\phi_0)^{l-1}|P_2;k>\;.
\label{form1}
\ee
We can write $f_k^{l-1}(P_1,P_2)=<P_1|(\hat{\phi}-\phi_0)^{l-1}|P_2;k>$
as in eq.(\ref{comple1})
\bqn
f_k^{l-1}(P_1,P_2)&=&
\sum_{m_i}
<P_1|(\hat{\phi}-\phi_0)|m_1>
<m_1|(\hat{\phi}-\phi_0)|m_2>
<m_2|...\nonumber\\
& &~~~~~~~~~~~~~...|m_{l-2}><m_{l-2}|(\hat{\phi}-\phi_0)|P_2;k>
\;.
\label{comple3}
\eqn
The left hand side of eq.(\ref{form1}) is of the order $O(\alpha^0)$
then $f_k^{l-1}(P_1,P_2)$ need to be of the order $O(\alpha^{2-l})$.
If we retain in eq.(\ref{comple3}) the sum only over soliton states
it can be seen that $f_k^{l-1}(P_1,P_2)$ is of required order
$O(\alpha^{2-l})$. But if we keep in the sum over $m_{l-2}$
soliton one meson states (in the other $m_i$'s we sum over soliton 
states only) using 
\be
<P_{l-2};k_{l-2}|\hat{\phi}|P_2;k>=2\pi\delta(k_{l-2}-
k)<P_{l-2}|\hat{\phi}|P_2>+<P_{l-2};k_{l-2}|\hat{\phi}|P_2;k>_c
\ee
and retaining only the disconected term we can see that
also such term will be of the required $O(\alpha^{2-l})$. Nextly
we can retain in the sums over $m_{l-2}$ and $m_{l-3}$ one meson
soliton states and retaining as above only the disconected terms
it can be seen that also the term obtained will be of the order 
$O(\alpha^{2-l})$. Next we retain in the sum over $m_{l-2}$, $m_{l-3}$
and $m_{l-4}$ soliton one meson states, next in $m_{l-2}$, $m_{l-3}$
$m_{l-4}$ and $m_{l-5}$ and so on. All these terms will be of the order
$O(\alpha^{2-l})$. We see that there is one term with origin 
in the sum over soliton states, and $(l-2)$ terms from summing
over soliton one meson states, that is we have $(l-1)$ terms
of the order $O(\alpha^{2-l})$. At this order still $E=M$. And
using a Fourier transformed function $\phi_k(x)$ 
\be
f_k(P_2,P_1)=\int dx \exp[i(P_2-P_1)x]\phi_k(x)
\ee
we can prove that all the $(l-1)$ terms give the same contribution in
eq.(\ref{form1}). Then, we obtain the following equation for $\phi_k(x)$
\bqn
\left[-\frac{d^2}{dx^2}-\omega^2(k)\right]\phi_k(x)&=&
-\sum_{l=2}^{\infty}l(l-1)a_l\alpha^{l-2}(\phi_c(x)-\phi_0)^{l-2}\phi_k(x)
\nonumber\\
&=&-U''(\phi_c(x))\phi_k(x)\;,
\eqn
that is, we obtain a Schrodinger like equation for $\phi_k(x)$
\be
\left[-\frac{d^2}{dx^2}+U''(\phi_c(x))\right]\phi_k(x)=\omega^2(k)\phi_k(x).
\label{diff}
\ee
Of course eq.(\ref{diff}) has discrete and continuum eigenvalues $\omega^2(k)$.
The discrete eigenvalues can be interpreted as excited
states of the soliton states. The continuum eigenvalues are 
$\omega(k)=\sqrt{k^2+m^2}$ (where $m$ is the mass of the mesons) and the eigevalues
asymptoticaly behave like $e^{ikx}$, {\it i.e}, meson plane waves. That is,
at this order the scattering of mesons by solitons are described effectively
by the scattering problem given by eq.(\ref{diff}). It is easy to show that
$d\phi_c/dx$ is the eigenfuntion of (\ref{diff}) with eigenvalue zero. It was shown
\cite{jackiw} that these eigenfuntions can be disregarded from the complete
basis $\phi_k(x)$ consistently with Poincare invariance and the conmmutation
relation between the field operator and its canonically conjugate momentum. The
zero mode solution of eq.(\ref{diff}) has no physical meaning, it gives only the
first corrections to the movement of the soliton. To show the above one
needs to write $\phi_k(x)$ in terms of normalized eigenfuntions $\psi_k(x)$,
as $\phi_k(x)=\psi_k(x)/\sqrt{2\omega(k)}$ and then we have for $f_k(P_2,P_1)$
\be
f_k(P_2,P_1)=\int dx \exp[i(P_2-P_1)x]\frac{\psi_k(x)}{\sqrt{2\omega(k)}}
\label{mode3}
\ee
where now we understand that the zero modes are excluded. Using eq.(\ref{mode3})
in expression (\ref{next}) one can prove that $f(P',P)$ is given by 
eq.(\ref{fourier}) but now with $\phi(x)$ given by
\be
\phi''(x)=U'(\phi)+\frac{1}{2}G(x,x)U'''(\phi)\;,
\label{corfor}
\ee
where 
\be
G(x,y)=\sum~'\frac{\psi_k^{\star}(x)\psi_k(y)}{2\omega(k)}\;,
\label{green}
\ee
where the prime in the sum indicate that the zero mode is excluded.
Also we can compute the mass of the soliton to the next order retaining
in (\ref{sumad}) the contribution of soliton one meson states, obtaining
\be
M=H[\phi_c]+\frac{1}{2}\sum_k\omega(k)\;.
\label{massq}
\ee

\section{The scattering of mesons by solitons}

\begin{figure}
\centerline {\epsfxsize=3.2in\epsfysize=1.8in\epsffile{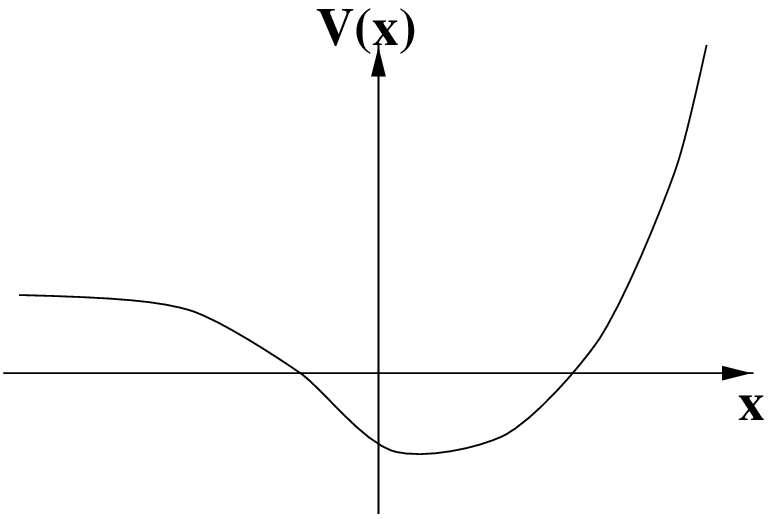} }
\caption{The Schrodinger potential for model I}
\begin{picture}(10,10)
\end{picture}
\label{morsev}
\end{figure}

As stated above to the scattering of mesons by solitons we have to
solve the one dimensional scattering problem for the Schrodinger equation
given by eq.(\ref{diff}). For the model I the eq.(\ref{diff}) is given by
\be
\left[-\frac{d^2}{dx^2}+m^2(e^{\pm2mx}-
3e^{\pm mx}+1)\right]\phi_k(x)=\omega^2(k)\phi_k(x)\;,
\label{diff1}
\ee
where the signs $\pm$ are respectively for the soliton anti-soliton sectors.
Here we solve explicitly the $+$ case. In fig.(\ref{morsev}) we show this potential for the $+$ sign. Solving 
eq.(\ref{diff1}) one finds that there is only one discrete 
eigenfuntion \cite{landau}
\be
\psi_0(x)=2\sqrt{m}e^{mx}\exp(-e^{mx})\;,
\label{zerom}
\ee
with zero eigenvalue $\omega^2_0=0$, and continuous eigenfuntions with eigenvalues going from $m^2$ to infinity. As we stated in preceding 
section the mode
zero has no physical interpretation. Because there is no more discrete
eigenfuntions, in this case the soliton state can not exist in a excited
state. The continuum eigenfuntions are interpreted as the scattering of 
mesons by the soliton. One can see from the potential profile showed in
fig.(\ref{morsev}) that mesons coming from $x=-\infty$ can not reach $x=\infty$,
they are absolutely reflected by the soliton going back again to $x=-\infty$.
Then one can espect to find for the reflection and transmitions
amplitudes $R=e^{i\delta(k)}$ and $T=0$ respectively, with $\delta(k)$ a real function.
In order to find $\delta$ we have to solve eq.(\ref{diff1}) with the
following boundary conditions
\be
\psi_k(x\rightarrow-\infty)=Ae^{ ikx}+Be^{- ikx}\;,
\label{asim1}
\ee
and
\be
\psi_k(x\rightarrow\infty)=0\;.
\label{asim2}
\ee
To compute $R$ we have to compute the relation between the values $A$ and $B$. For this end
we make the following change of variables
\be
\xi=2e^{mx},~~~\psi_k(x)=e^{-\xi/2}\xi^{ik/m}\chi(\xi)\;,
\label{chan}
\ee
in eq.(\ref{diff1}) and we obtain the following equation for $\chi(\xi)$
\be
\xi\chi''(\xi)+(2ik/m+1-\xi)\chi'(\xi)+(1-ik/m)\chi(\xi)=0\;,
\label{hyper}
\ee
a confluent hypergeometric equation, whose solution is given in terms of
hypergeometric functions\cite{landau}
\be
\chi(\xi)=c_1F(ik/m-1,2ik/m+1,\xi)+c_2\xi^{-2ik/m}F(-ik/m-1,1-2ik/m,\xi)\;.
\label{solhy}
\ee
Using eq.(\ref{solhy}) in eq.(\ref{chan}) we obtain for $\psi_k(x)$
\be
\psi_k(x)=e^{-\xi/2}\left[c_1\xi^{ik/m}F(ik/m-1,2ik/m+1,\xi)+
c_2\xi^{-ik/m}F(-ik/m-1,1-2ik/m,\xi)\right]\;,
\label{solum}
\ee
where we can see that for $\xi\rightarrow 0$ ($x \rightarrow-\infty$)
this solution behaves like eq.(\ref{asim1}). Using the asymptotic value for
$F(\alpha,\gamma,\xi)$ \cite{abramovitz}
\be
F(\alpha,\gamma,\xi\rightarrow\infty)=\frac{\Gamma(\gamma)}{\Gamma(\alpha)}
e^{\xi}\xi^{\alpha-\gamma}\;,
\label{asym}
\ee
it is easy to see that solution given by eq.(\ref{solum}) behaves as
eq.(\ref{asim2}) only if
\be
c_1=\frac{\Gamma(ik/m-1)\Gamma(-2ik/m)}{\Gamma(-1-ik/m)\Gamma(2ik/m)}c_2\;.
\label{coefs}
\ee
Now using eq.(\ref{coefs}) in eq.(\ref{solum}) and taking the limit
$x\rightarrow-\infty$ we obtain
\be
\psi_k(x\rightarrow-\infty)=c_2\left[\frac{\Gamma(ik/m-1)\Gamma(-2ik/m)}{\Gamma(-1-ik/m)\Gamma(2ik/m)}2^{ik/m}e^{ikx}+2^{-ik/m}e^{-ikx}\right]
\ee
from wich we obtain for the reflection amplitude:
\be
R(k)=B/A=2^{-2ik/m}\frac{\Gamma(-1-ik/m)\Gamma(2ik/m)}
{\Gamma(-1+ik/m)\Gamma(-2ik/m)}\;.
\label{refle}
\ee
And $\delta(k)$ is obtained as the argument of the above expression,
\be
\delta(k)=Arg\left(2^{-2ik/m}\frac{\Gamma(-1-ik/m)\Gamma(2ik/m)}
{\Gamma(-1+ik/m)\Gamma(-2ik/m)}\right)\;.
\ee
Note that as expected $|R(k)|=1$.
The case $(-)$ in eq.(\ref{diff1}) follows identicaly as the above case. The mesons
caming from $x\rightarrow\infty$ never reach $x\rightarrow-\infty$, they
are totally reflected by the soliton back again to $x\rightarrow\infty$. The
reflection amplitude is given as in the $(+)$ case by eq.(\ref{refle}).

\begin{figure}
\centerline {\epsfxsize=3.2in\epsfysize=1.8in\epsffile{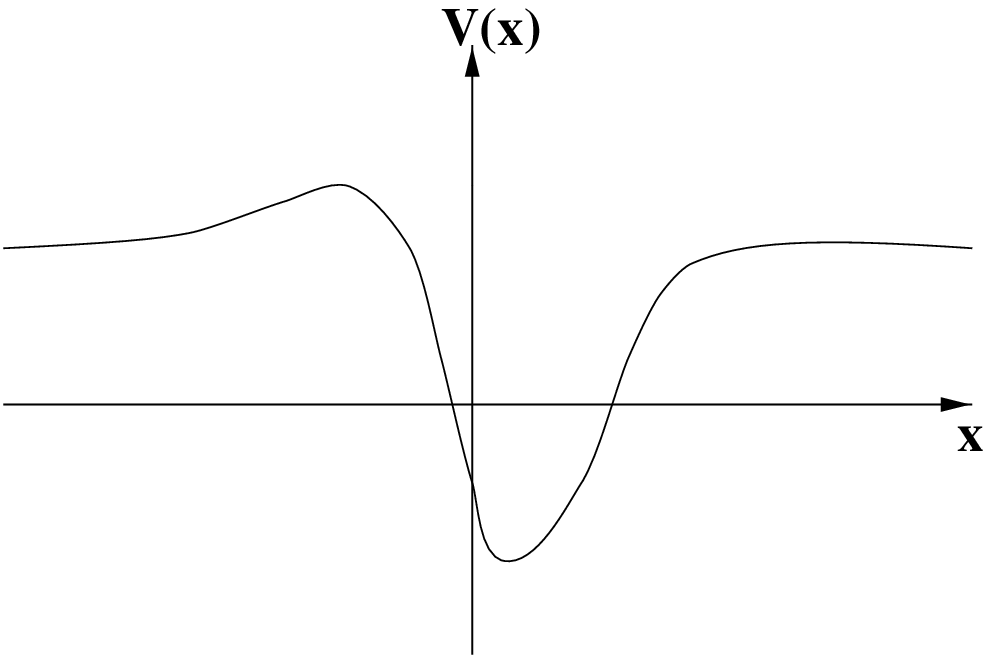} }
\caption{The Schrodinger potential for model II}
\begin{picture}(10,10)
\end{picture}
\label{scarfv}
\end{figure}

For model II the Schrodinger equation (\ref{diff}) is given by
\be
\left[-\frac{d^2}{dx^2}+m^2\left(1+\frac{(B^2-2)}{\cosh^2(mx)}
\mp 3B\frac{\tanh(mx)}{\cosh(mx)}\right)\right]\phi_k(x)=
\omega(k)^2\phi_k(x)\;,
\label{diff2}
\ee
where the signs ($\mp$) are related to the soliton and antisoliton sectors respectively. We show this potential in fig.(\ref{scarfv}) for the 
$-$ case.This equation have one discrete eigenvalue\cite{cooper}
\be
\psi_0(x)=N\frac{\exp(\pm 2B\tan^{-1}(\sinh(mx)))}{\cosh(mx)}\;,
\label{zerosc}
\ee
where $N$ is a normalization factor, and the continuum eigenfuntions with eigenvalues given by $\omega^2(k)=k^2+m^2$.
As in the model I there are not excited soliton states. From fig.(\ref{scarfv})
one can expect the following behaviour of  the scattering of mesons by the soliton (anti-soliton):
Mesons with momenta $k$ coming from $x=-\infty$   are scattered by the
soliton (anti-soliton) in mesons with momenta $k$ that reach $x=\infty$ (transmited mesons)
and in mesons with momenta $k$ that go back again to $x=-\infty$ (reflected mesons). To find the reflection and transmission amplitudes we have
to solve eq.(\ref{diff2}) with the adequate boundary conditions. The
solutions of eq.(\ref{diff2}) can be written in terms of hypergeometric 
functions and using the asymptotic properties of these hypergeometric
functions one can find easily the following for the reflection and 
transmision amplitudes\cite{khare}
\be
R(k)=\pm T(k)\frac{\sinh(\pi B)}{\cosh(\pi k/m)}\;,
\label{refles}
\ee
and
\be
T(k)=\frac{\Gamma(-1-ik/m)\Gamma(2-ik/m)\Gamma(\frac{1}{2}\mp iB-
ik/m)\Gamma(\frac{1}{2}\pm iB-ik/m)}
{\Gamma(-ik/m)\Gamma(1-ik/m)\Gamma^2(\frac{1}{2}-ik/m)}\;.
\label{trasms}
\ee
Note that $T(k)$ is the same for the scattering of mesons by the soliton 
and anti-soliton states. $R(k)$ only differ by a phase equal to $2\pi$, 
{\it i.e}, the scattering of mesons coming from $x=-\infty$ by the
soliton state is the same as the scattering of mesons coming from
$x=\infty$ by the anti-soliton state.  One can also verify that $(|R(k)|^2+|T(k)|^2)=1$.

\section{Quantum corrections for the soliton mass}
The first quantum correction for the soliton mass is given by the sum in eq.(\ref{massq})
over all the eigenvalues of the Schrodinger equation.
Such expression is divergent and if we subtract the zero point energy associated
with one of the trivial vacua the divergence is rendered logarithmic. Then we need
to renormalize such expression in order to obtain a finite expression. But since
we are in $1+1$ it is well known that all divergences can be eliminated by normal
ordering the field operator. Using this fact Cahill {\it et al} in reference \cite{cahill}
(see also \cite{tom}) found an expression free of divergences for the first quantum correction for the soliton mass. Expressed in terms of the normalized eigenfuntions
of the Schrodinger equation this finite expression is given by\cite{tom}
\be
\delta M=-\frac{1}{4}\sum_n\int_{-\infty}^{\infty}dq(\omega_n^2\omega_q^{-1}-2\omega_n
+\omega_q)|\widetilde{\psi}_n(q)|^2\;,
\label{finesum}
\ee
where
\be
\widetilde{\psi}_n(q)=\frac{1}{\sqrt{2\pi}}\int_{-\infty}^{\infty}dx
\exp(iqx)\psi_n(x)\;.
\label{ftr}
\ee
and $\omega_q=\sqrt{q^2+m^2}$. Although in both models I and II we know explicitly all the
eigenvalues and eigenfuntions it is still very hard to evaluate the expression given
by eq.(\ref{finesum}). But for model II there is a situation in which one can
obtain $\delta M$ aproximatelly. This situation is when $B$ is very small. In this
case from eq.(\ref{refles}) we see that $R\approx 0$, {\it i.e}, the Schrodinger equation
associated to the model II is approximatelly reflectionless. And as was shown in
ref.\cite{cahill} for reflectionless potentials, $\delta M$ given by eq.(\ref{finesum})
can be expressed in terms of the discrete eigenvalues only, {\it i.e},
\be
\delta M=-\frac{m}{\pi}\sum_{i}(\sin\theta_i-\theta_i\cos\theta_i)\;,
\label{cah}
\ee
where $\theta_i=\arccos(\omega_i/m)$. Since For model II there is only
one (zero) discrete eigenvalue we obtain $\delta M=-\pi/m$. Then taking
in equation (\ref{enrc2}) $I(B)\approx 2$ we can write
for $B$ small the following expression for the soliton masses asociated to 
model II
\be
M=\frac{18m^5B^2e^{2Bn\pi}}{\alpha^2}-\frac{m}{\pi}\;.
\label{masdos}
\ee
Since the above expression is valid for $B$ small one would naively discard 
the first term. This can not be made since $\alpha$ is a small parameter.

\section{Conclusions}
In this paper we have analyzed the quantum properties of solitons
in (1+1) field theory models with density potential given by equations
(\ref{new1}) and (\ref{new2}) that we called models I and II respectively.
Also we analyzed the scattering of mesons by solitons in both models at
$\hbar$ order. For model II we computed aproximatelly the first quantum 
corrections for the mass of the soliton quantum state. It is interesting
to remark that in model II the perturbative sector around the vacuum for
large $n$ is a free theory (see eq.(\ref{taylor2})) but the solitonic sector
is not. The mesons are scattered by the quantum soliton state. Then the
lesson that we have learned is that if a theory is free in one sector
there can be other sector(s) where it is not.

\vspace{0.5cm}

\begin{center}
{\large \bf Acknowledgements}
\end{center}

GHF and NFS are partially supported by Conselho Nacional de
Desenvolvimento Cient\'{\i}fico e Tecnol\'ogico - CNPq (Brazil).

\newpage

\end{document}